# Optimization of the n-dimensional sliding window inter-channel correlation algorithm for multi-core architecture

*Alexey Poyda[1] and Mikhail Zhizhin[2,3]*

[1]*National Research Center "Kurchatov Institute", Moscow, Russia*
[2]*Space Research Institute of the Russian Academy of Sciences, Moscow, Russia*
[3]*CIRES, University of Colorado, Boulder, Colorado, U.S.A.*

Calculating the correlation in a sliding window is a common method of statistical evaluation of the interconnect between two sets of data. And although the calculation of a single correlation coefficient is not resource-intensive and algorithmically complex, sequential computation in a large number of windows on large data sets can take quite a long time. In this case, each value in the data, falling into different windows, will be processed many times, increasing the complexity of the algorithm and the processing time. We took this fact into account and optimized the correlation calculation in the sliding window, reducing the number of operations in the overlapping area of the windows. In addition, we developed a parallel version of the optimized algorithm for the GPU architecture. Experimental studies have shown that for a 7x7 correlation window sliding in one pixel increments, we were able to accelerate the processing of an 12 MPixel image pixels on the GPU by about 60 times compared to the serial version running on the CPU. The article presents an optimized version of the algorithm, a scheme for its parallelization, as well as the results of experimental studies.

**Introduction**

Correlation in a sliding (or rolling) window is a very common method of statistical evaluation of synchronisation of two data sets used in image processing [1], signal analysis [2], medicine [3], geophysics [4], etc. The functions for calculating parameters in a sliding window (sum, arithmetic mean, correlation, etc.) are included in many software packages (for example, MATLAB, R, S-PLUS) and are actively used in the analysis of time series in financial statistics [5].

In this paper, we consider the Pearson linear correlation calculated in a sliding window between sets of spatial data on a regular grid, preferably between two images. The applied problem concerns the automatic search of fishing boat lights on multispectral satellite images obtained at night time [6]. The task is to filter the false detections caused by the reflection of moonlight from the clouds by calculating the correlation in the sliding window between the data in the visible and infrared channels. In the presence of clouds, the correlation coefficient should be close to -1, since the values in the visible channel increase due to the reflection of moonlight from the clouds, while the values in the infrared channel decrease due to the lower cloud temperature as compared to the ground surface.

The basic algorithm can be represented as a synchronous passage of two sets of data by a sliding window of a given size with a given step, and calculating the Pearson linear correlation between the values of the investigated sets in this window. In this case, the values in n-dimensional sliding windows are drawn into one-dimensional vectors.

To calculate the correlation in each window, the classical Pearson linear correlation formula is used:

$$c = \frac{\sum_{i=1}^{n}(x_i-\bar{x})(y_i-\bar{y})}{\sqrt{\sum_{i=1}^{n}(x_i-\bar{x})^2 \sum_{i=1}^{n}(y_i-\bar{y})^2}},$$

where $c$ is Pearson linear correlation between the $x$ and $y$ datasets;
$x_i$ and $y_i$ are the individual sample values;
$\bar{x}$ and $\bar{y}$ are the arithmetic means for the $x$ and $y$ datasets;
$n$ is the sample size.

Calculating the correlation in a sliding window using the classical formula leads to significant overhead, as illustrated by the example of processing 2D images, shown in Figure 1.

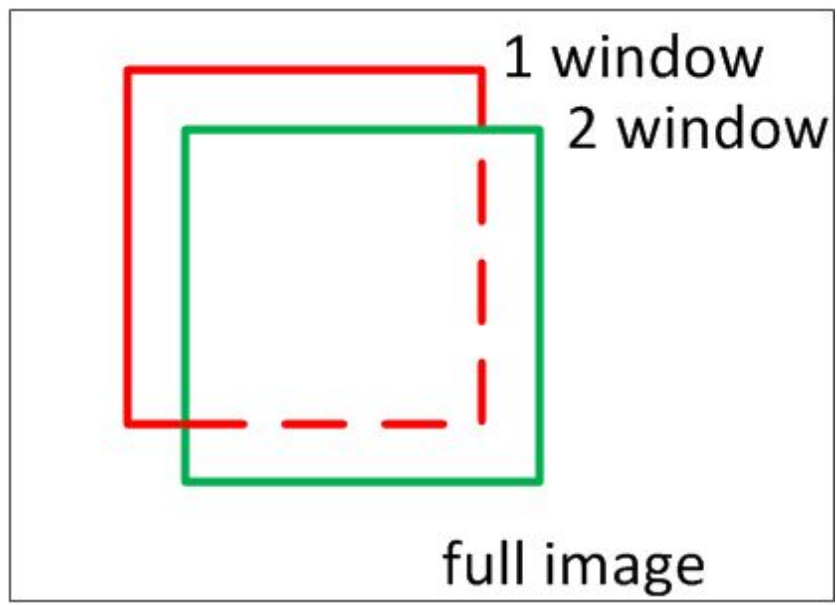

**Figure 1 - Calculation of the correlation in a sliding window using the classical formula using the example of two-dimensional images**

Windows 1 and 2 have a large overlap. When calculating the correlation in each window, the elements that belong to both windows will be processed twice. If the step of the sliding window is small enough, for example, one pixel, and the window size itself is large, then each element of the image will be processed as many times as the windows cover it.

In the work presented, we optimized the basic version of the algorithm for interchannel correlation in a sliding window, taking into account the modern trend to multi-core architecture, and have tested the optimized version on graphic processes.

**1. Optimized version of the algorithm**

To reduce the overhead caused by multiple processing of the same element, we used a modified correlation calculation formula that is often used to reduce the number of computational operations associated with subtracting from each arithmetic mean value:

$$c = \frac{n \sum x_i y_i - \sum x_i \sum y_i}{\sqrt{n \sum x_i^2 - (\sum x_i)^2} \sqrt{n \sum y_i^2 - (\sum y_i)^2}}$$

An advantage of this formula for the problem studied in this paper is the expression of the correlation value through the sums of the quantities $x$, $y$, $xy$, $x^2$, $y^2$. Such sums for one-dimensional series can be obtained by a linear algorithm for calculating the sum in the sliding window of the movements (similar to the moving average of the *rollmean* R function in the *zoo* library[1]). To move from a one-dimensional series to a two-dimensional image matrix, we consistently apply the "sliding summation" first in one dimension, and then in the second dimension (Figure 2).

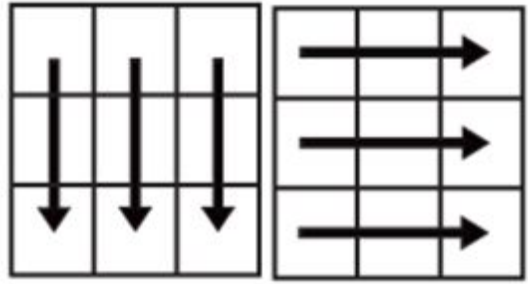

**Figure 2 - The scheme of calculating the sum in a sliding window for the two-dimensional case**

As a result of the superposition of two one-dimensional summations with a window of length k, we obtain a matrix of the same size as the original, but at each position of which there will be a sum of elements of this matrix taken in a window of size k * k centered at a given point (provided that the one-dimensional summation also puts the result in the center of the window).

---

[1] https://CRAN.R-project.org/package=zoo

Thus, the optimized algorithm is as follows:
1. Element multiplication of matrices x and y computes the values of matrices xy, $x^2$, $y^2$.
2. For each matrix x, y, xy, $x^2$, $y^2$, a summation is made with a sliding window of length k over all rows (the results are denoted by Ax, Ay, Axy, $Ax^2$, $Ay^2$).
3. For each matrix Ax, Ay, Axy, $Ax^2$, $Ay^2$, a summation is made with a sliding window of length k over all columns; the results are denoted by Sx, Sy, Sxy, $Sx^2$, $Sy^2$ (the sum matrix in the windows centered on this element).
4. For each element of the resulting matrix, we calculate the correlation by the formula 1.

In order to generalize the above algorithm to the n-dimensional case, it is necessary to repeat the summation with the sliding window (steps 2-3) n times, one for each measurement.

It is worth noting that optimization can be done not only with the help of a sliding window, but also with the help of cumulative sums [7]. We have tested this approach, but in our application the cumulative sums give an increase in the execution time in comparison with the algorithm proposed in this section. Theoretical calculations (which we do not show here to save space) confirm the increase in the number of operations using cumulative sums compared to the summation in a sliding window.

## 2. Parallelizing an optimized version of the algorithm

The algorithm described in the previous section can be efficiently parallelized for the case of a multi-core architecture, for example, the GPU:
1. In step 1, each thread processes elements from a small neighborhood of the source data.
2. In step 2, each thread processes one or more rows, depending on the size of the data and the degree of parallelism.
3. In step 3, each thread processes one or more columns, depending on the size of the data and the degree of parallelism.
4. In step 4, each thread processes elements from a small neighborhood of the source data.
5. For the n-dimensional case, parallelization is performed analogously.

## 3. Implementation of algorithms and testing

The algorithms described in the previous sections were implemented in software. The basic and optimized versions were implemented in the programming language MATLAB. The parallel optimized version is implemented in C ++ programming language for CUDA libraries. In the software implementation, the correlation in the sliding window between two two-dimensional data sets is calculated. Step of the window is one cell. Both sets of data must be the same size.

The result set is the same size as the arguments. For each pair of windows, the correlation result is written to the middle cell. Data may contain missing values that are identified by a number less than or equal to -999. If at least one missing value occurs in the correlated windows, then the resulting value for the two given windows is set to -2. Values along the edges of the resulting matrix, for which the correlation windows corresponding to them exceed the boundaries of the matrix-arguments, are set equal to -2.

Testing and comparison of software implementations was carried out on two satellite images, reprojected to a regular geographic grid. In particular, we used images in the visible and infrared channels obtained by the VIIRS sensor installed on the Suomi NPP satellite. The size of the images is 4064x3072 pixels. The size of the correlation window is 7x7 pixels.

The MATLAB version was tested on a workstation with the following characteristics:
- Processor: 2 x Intel Xeon E5640 (8 cores, 2.66 GHz)
- RAM: 8 GB
- Operating system: Windows 7

The GPU version was tested on a computer with the following characteristics:
- processor: 2 x Intel Xeon E5-2650 v2 (8 cores, 2.60 GHz)
- RAM: 128 GB
- graphics card: 2 NVIDIA Tesla K80 cards
- operating system: Ubuntu 14.04.3

Obviously, the overall performance of a computer with a GPU is significantly higher than that of a computer with a CPU, so it would seem that the comparison is incorrect. But the essence of the experiment is not to compare the specific performance of two different architectures, but to evaluate the effect and quality threading of a single-threaded algorithm, as well as the efficiency of running a parallel algorithm on a GPU architecture (the latter is not equivalent to the first one, since the algorithm can be paralleled well, but its implementation on the GPU architecture may not display a large increase in performance due to various technical limitations, for example, the cost of sending data can significantly exceed the cost of processing them.) If parallelization is effective, then the loading of the GPU in our test will be high, and the processing speed should be significantly reduced in comparison with the CPU, so as the overall performance of the GPU is higher. If parallelization is inefficient, then either we start with a large overhead for organizing parallelization on the GPU, or only a small fraction of the GPU will be downloaded der (and we know that one GPU core has suschestenno lower performance than a single CPU core so the GPU efficiency is achieved only through massive parallelism). In both cases, we will not get a significant performance gain on any GPU-device, and may even get the opposite effect - the slowdown of the algorithm.

The testing showed the following results:
- Basic version: about 42 seconds, about 1 second - data reading and about 41 seconds - the work of the correlation algorithm itself.
- Optimized version: about 3.3 seconds, about 1 second - data reading and about 2.3 seconds - the work of the correlation algorithm itself.
- Parallelized optimized version: about 13 seconds, of which: almost 11 seconds - initialization of the GPU, 1.5 seconds - reading the data and about 0.6 seconds - the work of the algorithm itself, including allocating memory to the GPU and transferring data to and from the device.

Thus, optimization accelerated the algorithm for the test case by more than 15 times, and parallelization and transfer to the GPU accelerated it by about 4 times.

If we make a theoretical calculation of the number of arithmetic operations when processing 2D data, we get:
- For the basic version: [2 * n * n {calculating two arithmetic means} + 2 * n * n {subtraction from each element of the arithmetic mean) + n * n {multiplication in the numerator} + n * n {addition in the numerator} + 2 * n * n {raising the elements to the second power in the denominator} + 2 * n * n {two summations in the denominator} + 2 {calculating the square root in the denominator and dividing}] * (A-n + 1) * (B-n +1) {number of windows} ≈ $10 * n^2 * A * B$, where n is the size of one dimension of the window, A and B are the dimensions of the two-dimensional data set being processed.
- For the optimized version: 3 * A * B {the elementwise calculation of the values xy, $x^2$, $y^2$} + 5 * SM + 3 * A * B {two element products and subtraction} + 2 * 3 * A * B {two calculations of the root expressions, each of which includes two element-wise multiplications and an element-by-element difference} + 3 * AB {two element-wise square root calculations} + A * B {element division} = 16 * A * B + 5 * SM, where A and B are the sizes of the two- data, and SM - the number of operations necessary to calculate the two-dimensional summation with a sliding window. SM = A {number of lines} * [B {number of summations of elements in one line} + (Bn) {number of subtractions of elements in one line}] + B {number of columns} * [A {number of summations of elements in one column} + ( An) {the number of subtractions of elements in one line}] = A * [B + (Bn)] + B * [A + (An)] ≈ 4 * A * B. Altogether, for the optimized version of the algorithm we get 16 * A * B + 5 * SM = 36 * A * B arithmetic operations.

Of course, the above calculations can not accurately reflect the time relationship, since not all operations are performed the same time, and also because we did not take into account computer operations, for example, assigning values, but nevertheless these formulas can be used to approximate the complexity of algorithms. If we substitute in the above formulas the values used in the experiment, we get:
- for the basic version of the algorithm: $10 * n^2 * A * B = \{n = 7\} = 490 * A * B$;
- for the optimized version of the algorithm: $36 * A * B$.

And their ratio is $490/36 \approx 13.5$, which is pretty close to the ratio of the execution time of these algorithms, obtained during testing.

Analyzing the received formulas it is possible to draw a conclusion: the number of arithmetic operations in the basic version of the algorithm depends squarely on the window size, whereas in the optimized algorithm the amount of arithmetic operations does not depend on the window size at all.

Similar theoretical calculations can be made for a parallel version, but they will differ greatly from the practical implementation for the following reasons:
1. We can not in the theoretical calculations take into account the overhead associated with the initialization and support of one computational flow. Moreover, different computing architectures will require different overhead.
2. It is difficult for us to take into account the degree of parallelism in a particular architecture and the particular implementation. For example, not all parallel threads can run in parallel. For example, in the case of a GPU, a particular partition of the data space into blocks can greatly affect performance.
3. It is difficult to take into account the costs of data exchange. For example, again, in the case of a GPU, the compiler can optimize the program by placing data on registers or in local memory. And this leads to an increase or decrease in the operating time of the program.

The most critical issue of accuracy arises when all values of the correlation window are equal, in which case the denominator in the correlation calculation formula goes to zero, but even a small calculation error can lead to deviation of the calculated value from zero. As a result, the divide-by-zero error will not arise, and the resulting value will be incorrect. To avoid this situation, we propose to check all the elements for equality before carrying out the basic calculations.

Nevertheless, if the extreme values are not taken into account, then the error of the algorithms is rather small: if the results of the base version are taken as a reference result, then in our test for the optimized version the maximum error is less than $10^{-3}$.

**Conclusion**

The optimized version of the algorithm for calculating inter-channel correlation in a sliding is used in VIIRS boat detection algorithm [6] to filter false detections of fishing boat lights caused by the reflection of moonlight from clouds. Parallel implementation have reduced computational time by more than 15 times for a multicore CPU architecture, and 2 times more for a CUDA GPU. Initialization time of the GPU is not taken into account, since initialization can be done in advance, when the program starts.